\documentclass[journal=jacsat,manuscript=article]{achemso}

\usepackage[version=3]{mhchem} 
\usepackage{multirow}

\author{Stephen Wu}
\affiliation[ISM]
{The Institute of Statistical Mathematics, Research Organization of Information and Systems, Tachikawa, Tokyo 190-8562, Japan}
\alsoaffiliation[SOKENDAI]
{The Graduate University for Advanced Studies, SOKENDAI, Tachikawa, Tokyo 190-8562, Japan}
\email{stewu@ism.ac.jp}

\author{Hironao Yamada}
\affiliation[ISM]
{The Institute of Statistical Mathematics, Research Organization of Information and Systems, Tachikawa, Tokyo 190-8562, Japan}
\affiliation[TUPLS]
{School of Pharmacy, Tokyo University of Pharmacy and Life Sciences, Hachioji, Tokyo 192-0392, Japan}

\author{Yoshihiro Hayashi}
\affiliation[ISM]
{The Institute of Statistical Mathematics, Research Organization of Information and Systems, Tachikawa, Tokyo 190-8562, Japan}

\author{Massimiliano Zamengo}
\affiliation[TITECH]
{School of Materials and Chemical Technology, Tokyo Institute of Technology, Meguro, Tokyo 152-8550, Japan}

\author{Ryo Yoshida}
\affiliation[ISM]
{The Institute of Statistical Mathematics, Research Organization of Information and Systems, Tachikawa, Tokyo 190-8562, Japan}
\alsoaffiliation[SOKENDAI]
{The Graduate University for Advanced Studies, SOKENDAI, Tachikawa, Tokyo 190-8562, Japan}
\alsoaffiliation[NIMS]
{National Institute for Materials Science, Tsukuba, Ibaraki 305-0047, Japan}

\title[An \textsf{achemso} demo]
  {Potentials and challenges of polymer informatics: exploiting machine learning for polymer design}

\begin{document}







\begin{abstract}
  There has been rapidly growing demand of polymeric materials coming from different aspects of modern life because of the highly diverse physical and chemical properties of polymers. Polymer informatics is an interdisciplinary research field of polymer science, computer science, information science and machine learning that serves as a platform to exploit existing polymer data for efficient design of functional polymers. Despite many potential benefits of employing a data-driven approach to polymer design, there has been notable challenges of the development of polymer informatics attributed to the complex hierarchical structures of polymers, such as the lack of open databases and unified structural representation. In this study, we review and discuss the applications of machine learning on different aspects of the polymer design process through four perspectives: polymer databases, representation (descriptor) of polymers, predictive models for polymer properties, and polymer design strategy. We hope that this paper can serve as an entry point for researchers interested in the field of polymer informatics.
\end{abstract}

\section{Introduction}
Polymers are one of the most important classes of material in modern society, as its applications range from the plastic bags and bottles used in daily life to a variety of electronics, and even structural components in the aerospace industry. A polymer is a material made of a collection of chains that are built by connecting many repeated units, called monomers. These chains can form diverse structures that contribute to the highly diverse physical and chemical properties of different types of polymers. Some polymers can be consisting of more than one type of monomer to form even more complicated topological structures across different length scales. The research field of polymer science and engineering emerged to understand, control, and design novel polymers that can be used to satisfy the rapidly growing demand on highly functional materials coming from different aspects of modern life. While polymers can be categorized into natural or synthetic polymers, we will focus our discussion on the latter in this paper. 

Following the mainstream of materials science, polymer science has gone through multiple major paradigm shifts. The early studies of polymers flourished in the first half of the 20th century was closely associated with H. Staudinger who received the Nobel Prize in 1953\cite{polymer_history}. Initially, discovery of polymers was mainly based on an empirical approach, i.e., relying on many trial-and-error experiments. Accumulation of experimental experiences has led to developments of theoretical and simple statistical models for guiding the design of new polymers. These include many important work by the 1974 Nobel prize recipient P. J. Flory, the group contribution method and so on\cite{Flory1969, VKBook, BiceranoBook}. Following the rapid advances of computing power in the last few decades, computational methods has become one of the main tools to study properties of polymers\cite{polymer_MD}. 
Recent developments of simulation techniques for various types of polymers opened up opportunities to computationally study polymers across different length scales\cite{MDReview2009, Gartner:2019aa}. While generating experimental data of polymers is often costly and time-consuming, modern supercomputers provide new opportunities for building larger databases of polymers computationally. With the drastic expansion of data size in science, a data-driven approach for scientific discovery is said to be the 4th paradigm of science. Polymer informatics is an interdisciplinary research field of polymer science, computer science, information science and machine learning that serves as a platform to mine the precious polymer data for new knowledge. Yet, there has been notable challenges of the development of polymer informatics attributed to the complex hierarchical structures of polymers\cite{Audus:2017aa, kumar_li_jun_2019}.

\begin{figure}
  \includegraphics[trim={0 0 0 0},clip, width=14cm]{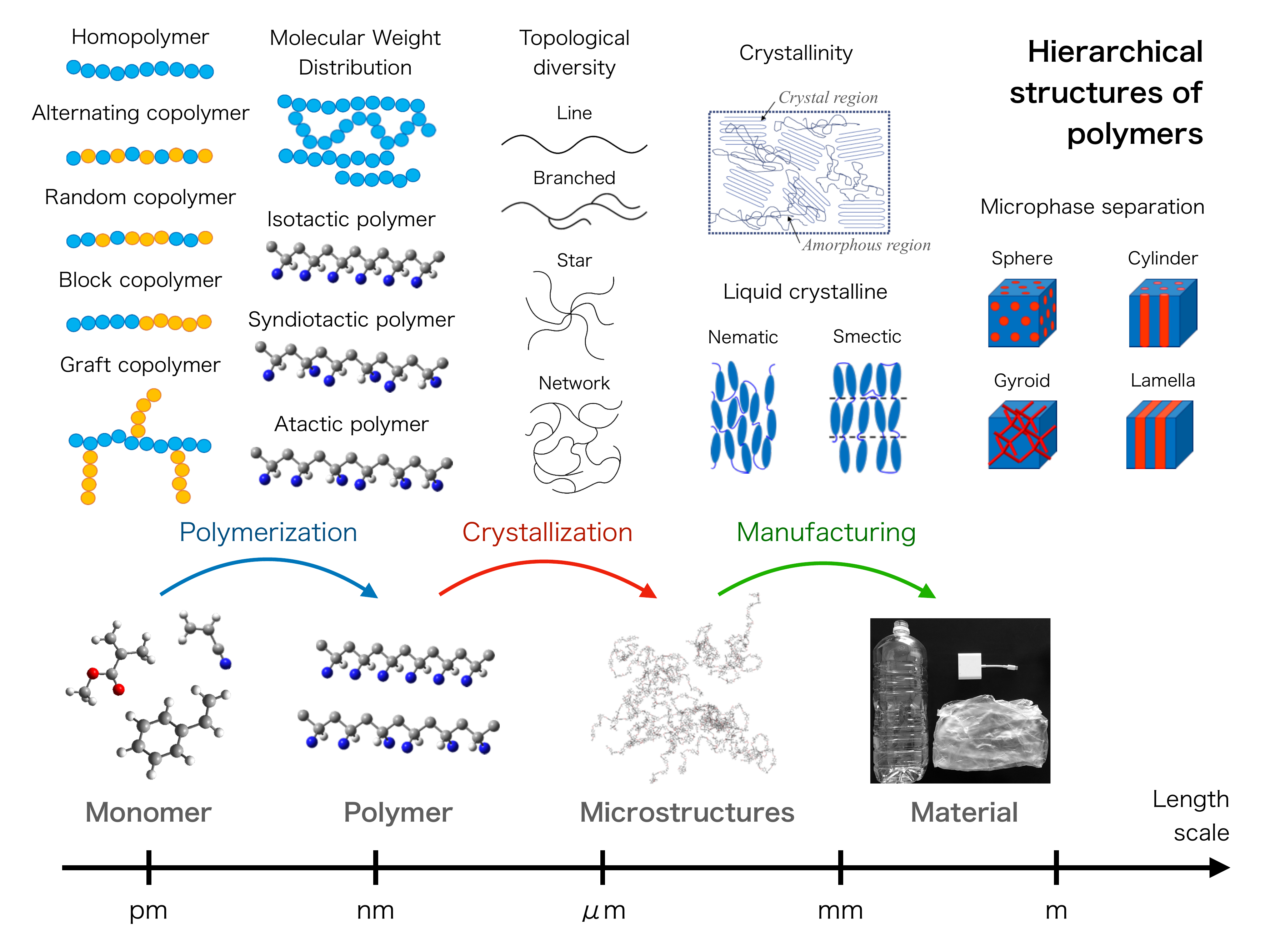}
  \caption{Overview of polymer design across different length scales.}
  \label{fig:polymer_production}
\end{figure}

Design of a polymer can be broken down into three parts corresponding to the three steps in the typical production process of polymers: design of monomers (polymerization), microstructures (crystallization), and material processing (manufacturing) (see Figure \ref{fig:polymer_production}). Monomers, the building blocks of polymers, contribute to the foundation of potential properties of the eventually produced polymers. While molecular size is one of the important factors that influences the properties of an organic material, the ``size effect" of polymers is not directly correlated with the size of the monomers because a large collection of small monomers (e.g., ethylene) can also build long polymer chains the same way large monomers do. Instead, various metrics based on the molecular weight distribution (MWD) of a polymer are often used as a reference to relate the ``size of polymer" to polymer properties. Different polymers built from the same monomer can have different MWD by controlling the polymerization process, which can lead to significantly different physical and chemical properties\cite{Imrie:1994aa, doi:10.1002/pen.760220402, Fetters:1994aa}. 
Furthermore, the collection of polymer chains can form very different crystal structures through a variety of crystallization processes, affecting the material properties of the resulting polymers. For example, \citeauthor{C8NR05407J} controlled the crystallinity and orientation of poly(3-hexylthiophene) molecules to optimize the performance of solar cells\cite{C8NR05407J}. Finally, the same type of polymer in the microscopic scale can undergo different manufacturing process, such as stretching or mixing additives, to further enhance or alter its properties in order to fulfil specific needs from a broad range of applications\cite{PolyethleneBK}. Ideally, the design space of polymers covers all parameters involved in the three production steps, for example, the molecule space of a single or multiple monomer(s) (namely homopolymers or copolymers, respectively), temperature and types of polymerization process, additives or fillers, molding methods, etc. In practice, we often focus on a subset of the parameters while keeping other fixed to reduce the enormous search space. For example, \citeauthor{Wu:2019aa} focused on the design of monomer that has a high probability of making liquid crystal polymers with high thermal conductivity after compressed to a polymer thin film\cite{Wu:2019aa}. 

\begin{figure}
  \includegraphics[trim={0 0 0 0},clip, width=16cm]{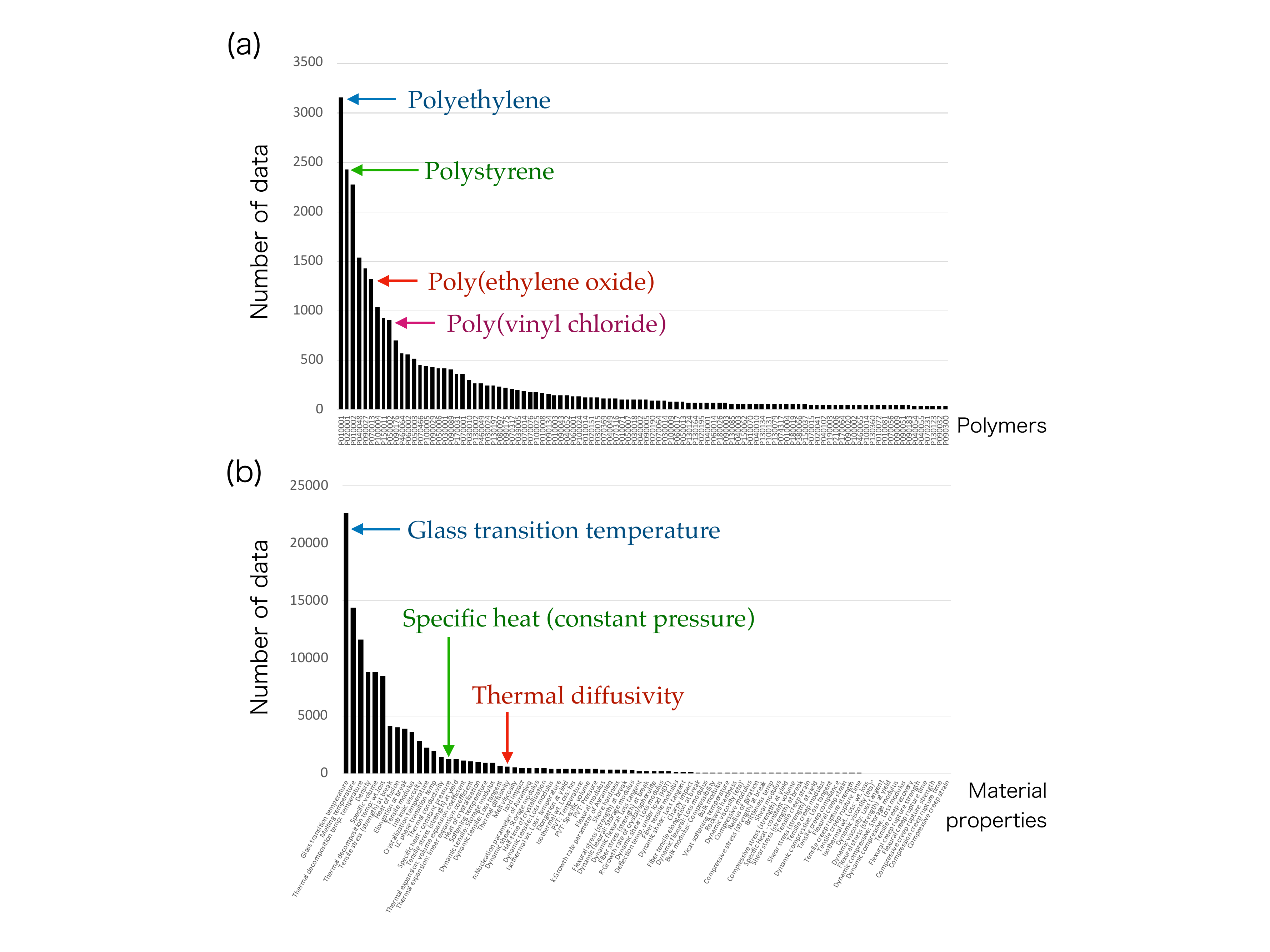}
  \caption{Statistics of 54151 data entries of 83 different polymer properties related to thermal properties recorded in PoLyInfo (extracted on April 2016). (a) Histogram of the number of data for the top 100 polymers with the most data is plotted in descending order. (b) Histogram of the number of data for 83 different polymer properties is plotted in descending order.}
  \label{fig:PI_data}
\end{figure}

A machine learning approach of polymers design on such a large design space requires a very large data set, in terms of both quantity and diversity. Unfortunately, openly available large polymer database is still limited\cite{Audus:2017aa} and historical data is often highly biased to a few types of polymers or polymer properties. For example, in PoLyInfo\cite{PoLyInfo, PoLyInfo:2011}, which is one of the largest database of polymers, around 30\% of data related to thermal properties are covered by only 10 different polymers and over 40\% of data is glass transition temperature (see Figure \ref{fig:PI_data}). 
In order to achieve a fully data-driven process of polymer design, continuous efforts have been made to bridge the demand of machine learning technologies and the current state of polymer informatics. In this paper, we review and discuss the applications of machine learning on different aspects of the polymer design process through four perspectives: polymer databases, representation (descriptor) of polymers, predictive models for polymer properties, and polymer design strategy. Illustrative examples are also given using the open-source materials informatics software, XenonPy\cite{XenonPy}. We hope that this paper can serve as an entry point for researchers interested in the field of polymer informatics, who may be coming from any of the scientific fields covered by polymer informatics.

\section{Machine learning in polymer informatics}

The goal of machine learning is to develop computer algorithms that can automatically improve their ability to solve a target problem by extracting information from past experience (training data). A basic implementation of this idea is to build a mapping $f$ from an input $x$ to an output $y$ when given some relevant data $D$. Here, $x$ is the representation (or descriptor) of a problem of interest, which we will discuss in detail in the next section. Depending on the choice of $f$, $y$ and $D$, machine learning is often categorized into:
\begin{itemize}
    \item \textbf{Supervised learning} ~~ Directly building $f$ that maps $x$ to the desired output of interest $y$ by learning from many examples of pairings between different $x$ and $y$ in $D$. Such data that contains pairs of $x$ and $y$ is called labelled data. Supervised machine learning is often subcategorized into regression or classification, where $y$ is either a continuous variable or a discrete variable, respectively. An example would be building a model to predict the glass transition temperature of homopolymers ($y$) from composition and bonding information of the corresponding monomers ($x$)\cite{Wu:2016aa, Kim:2018aa}.
    \item \textbf{Unsupervised learning} ~~ Learning the underlying structure of $x$ using unlabelled data, i.e., $D$ has no information about the output of interest $y$. Typical techniques of unsupervised machine learning are clustering and dimension reduction, where $x$ is mapped to some categorical labels or lower dimensional space, respectively. For example, \citeauthor{PhysRevE.99.043307} used these techniques to study phase transitions of polymer configurations\cite{PhysRevE.99.043307}. 
    Note that the learned mapping is not necessarily directly correlated with the targeted $y$ as such information is not included in $D$.
    \item \textbf{Reinforcement learning} ~~ Learning a strategy to achieve a certain goal in an interactive environment. Closely related to the problem of experimental design, the goal here is to learn $f$ that maps the current state $x$ to a possible action $y$. Successful training of $f$ through repeated engagement to the system can take the system closer to the final goal. Typically, a reward function is defined to quantify the progress of the system and the size of $D$ increases continuously during the interactive engagement process. An example would be developing optimal strategy to control MWD of a class of polymers\cite{C7ME00131B}.
\end{itemize}

Depending on the nature of application, available data and computing resources, it is important for researchers to frame their problems under an appropriate class of machine learning, to pick a suitable learning algorithm, and to represent the materials of interest using an effective descriptor. We attempt to provide useful hints from existing literature and our own experience in the following sections. 

\section{Database}

One of the most important components in machine learning is data. The quality and quantity of available data determine the scope of solvable problems in an application. A rule of thumb in machine learning is that purely data-driven models are not reliable when extrapolating. In other words, it is dangerous to trust the prediction of a model on a material that is not similar to the ones in the training data. Therefore, high quality large database for a diverse set of polymers is always of high demand in polymer informatics. Table \ref{tab:DB} shows a list of online databases that contain polymer data. There also exists large amount of publications on polymer technology or database that collects recipes of polymer synthesis (e.g., NIST Synthetic Polymer MALDI Recipes Database\cite{MALDI}). These types of information require extra processing effort before being useful for polymer informatics. 

\begin{table}
  \caption{List of online polymer databases. Numbers are extracted on September 25, 2020.}
  \label{tab:DB}
  \begin{tabular}{p{0.35\textwidth}p{0.55\textwidth}}
    \hline
    Name (link) & Descriptions  \\
    \hline
    PoLyInfo (polymer.nims.go.jp) & Polymer database supported by National Institute for Materials Science (NIMS) where data is mainly extracted from academic literature (covering 18,044 literature data). The database includes 367,711 property data points of various kinds of polymers built from 18,015 different monomers\cite{PoLyInfo, PoLyInfo:2011}. \\[0.3cm]
    Polymer Genome - Khazana (khazana.gatech.edu) & An open online platform that stores computational and experimental data from 24 publications. The database includes property data of 1,412 different polymers/organic materials and 2,657 different inorganic materials\cite{Huan:2016aa, Kim:2018aa}.  \\[0.3cm]
    Polymer Property Predictor and Database (pppdb.uchicago.edu) & Online polymer database maintained by CHiMaD that includes 263 and 212 data entries of Flory-Huggins $\chi$ value and glass transition temperature, respectively, extracted from the literature. \\[0.3cm]
    NanoMine ~~~~~~~~~~~~~~~~~~~~~~~~~~~~~~~~ (materialsmine.org/nm\#) & An open platform for data sharing that includes images of polymer microstructures and property data of polymers\cite{Zhao:2016aa, Zhao:2018aa}. \\[0.3cm]
    Cambridge Structural Database (www.ccdc.cam.ac.uk/structures) & Crystal structure database of organic and inorganic materials that includes more than 1,000,000 structures, where around 11\% is polymeric. \\[0.3cm]
    CROW (polymerdatabase.com) & An online data source that includes thermo-physical data of polymers. The source is either experimental data from the literature and/or calculated values from similarity analysis or quantitative structure property relationships. \\[0.3cm]
    Polymers: A Property Database (poly.chemnetbase.com) & Online database of various polymer properties used to support the book \textit{Polymers: A Property Database} by Wiley\cite{PDB_book}. \\[0.3cm]
    Citrination (citrination.com) & Materials informatics platform that includes publically available data of mechanical properties and solid surface energy of polymers. \\[0.3cm]
    CAMPUS (www.campusplastics.com) & Material property database of 9,236 commercial polymer grades. \\[0.3cm]
    Identify (www.netzsch-thermal-analysis.com) &
    Commercial software and database that includes differential scanning calorimetry curves for more than 600 commercial polymers. \\
    \hline
  \end{tabular}
\end{table}


Comparing to other research fields (e.g., image recognition) that benefit from modern machine learning technology, deep learning specifically, the number and size of open databases in polymer informatics are significantly smaller. We have accumulated a large amount of polymer data throughout the history of polymer science, but many historical data recorded in handbooks or publications are not well-organized, and a lot of the industry-owned data are not openly available. These issues have become the bottleneck for the development of polymer informatics\cite{Audus:2017aa}. With the advancements of computational simulation technologies and supercomputers, we expect an increasing interest and opportunity to build large scale computational databases of polymers. Meanwhile, technology of high-throughput experiments\cite{Oliver:2019aa} and the use of robotics combined with artificial intelligence\cite{Burger:2020aa} provide new opportunities to build experimental database of polymers efficiently. Making these databases open for research purposes will be the key to the success of polymer informatics.

\section{Descriptor}

The fundamental purpose of a material descriptor is to uniquely encode materials in a compact form in order to allow for efficient machine learning. This is particularly difficult for polymers due to their hierarchical structures\cite{HS_polymer}. An example of unique representation of polymers would be the Polymer Markup Language, which is designed to include complete information of a polymer ranging from compositional information to all the processing parameters\cite{Adams:2008aa}. While this representation may be suitable for building polymer databases, it is not compact enough to be used as input of a model for machine learning purposes. In fact, a good material descriptor should consider the tradeoff between ability to uniquely represent a material, easiness to obtain or calculate, and sensitivity to targeted application\cite{ZHOU20191017}. In other words, while possible, it is unlikely that a single descriptor can be used for all polymer applications. For example, to search for an efficient polyimide for controlled inkjet deposition, descriptor needs to be sensitive to specific microstructures formed by the polymer chains\cite{C5PY00622H}. On the other hand, designing polymers with certain phase transition properties will require a descriptor that identifies key atomic-level structures. Different level of ``fineness" of descriptor should be selected depending on the physical and chemical properties of interest to the target problem\cite{Ramprasad:2017aa}. One example that attempted to capture such hierarchy of descriptor in a Python package is the Materials knowledge systems\cite{Brough:2017aa}. Here, we introduce and compare a few descriptors that could be useful in polymer informatics.

Processing conditions of polymers, such as temperature, additives and solvents, can be directly included in a descriptor. Representations of the microstructure and molecular structure of polymer chains are less intuitive. Certain machine learning models in deep learning allow direct use of images as input, but often require a significant amount of data to train the models\cite{NIPS2012_4824}. Persistant homology is a technique to extract statistical features from topological structures\cite{Buchet2018}, but its application in polymer informatics is yet to be investigated. Graphical kernel is another option to numerically encode the polymer structures that can be represented in a simple graph\cite{JMLR:v11:vishwanathan10a}. However, finding an efficient graph representation for the complex polymer structures is challenging, especially because different types of polymer chains can form structures in different length scales. Many of the descriptors commonly used in polymer informatics focus on capturing the molecular structure of monomers composing the polymer chains. 

\begin{figure}
  \includegraphics[trim={0 0 0 0},clip, width=12cm]{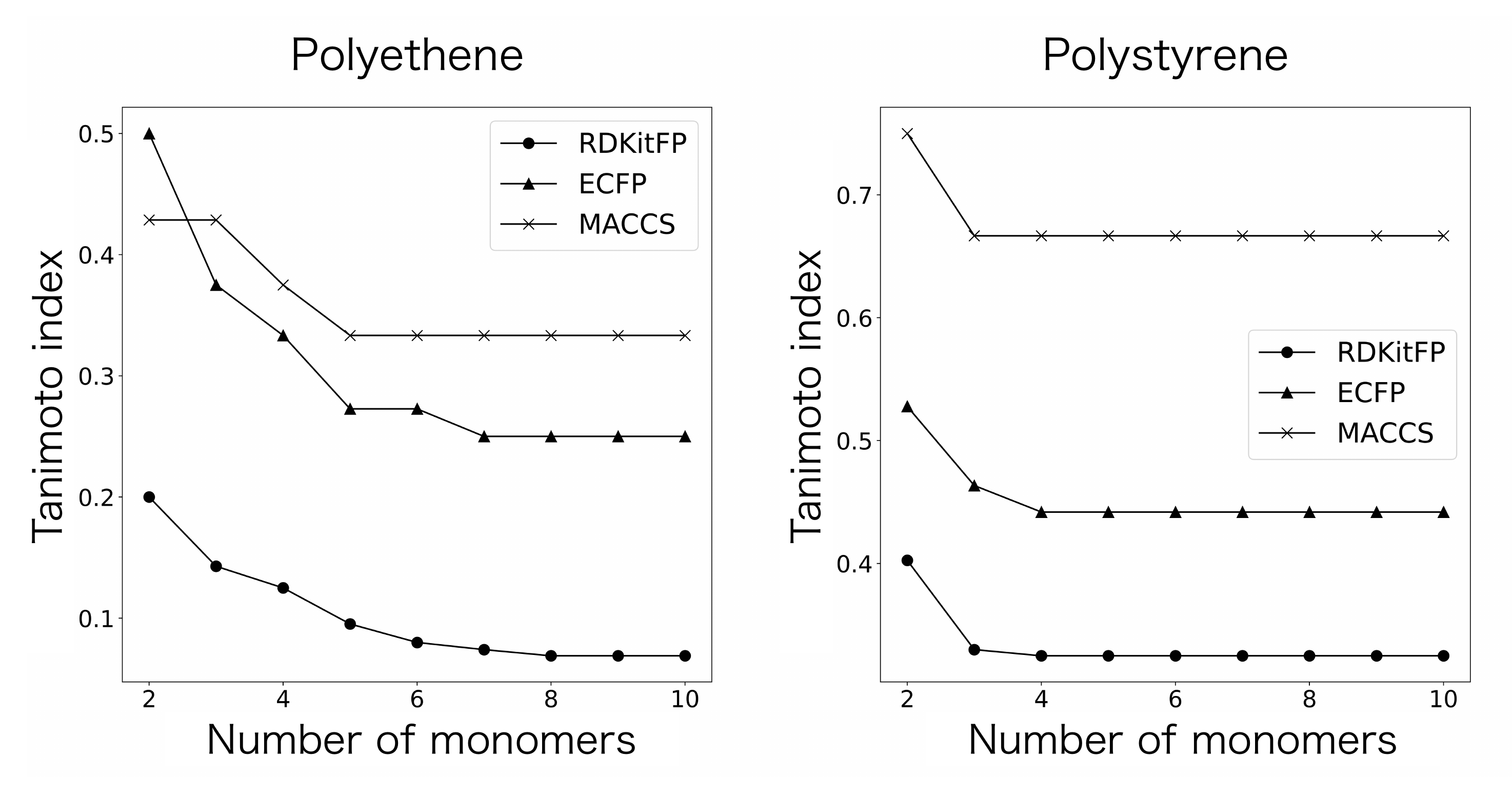}
  \caption{Convergence of the change of fingerprints for two polymers as a function of the number of monomers in the oligmer. The change is measured by Tanimoto index between the fingerprints of the monomer and the oligmer ($n$ from 2 to 10), which represents the similarity of the two fingerprints. Three RDKit fingerprints implemented in XenonPy are used: \textit{RDKitFP} denotes the standard fingerprint, \textit{ECFP} denotes the Morgan fingerprint, and \textit{MACCS} denotes the MACCS keys.}
  \label{fig:fingerprints}
\end{figure}

Material descriptors or fingerprints developed for small organic molecules are directly applicable to represent polymers based on their monomers. 
Modern machine learning algorithms also allow direct use of molecular graph or simplified molecular-input line-entry system (SMILES)\cite{SMILES:1988} strings as input of a model\cite{MICCIO2020122341}. Picking a descriptor suitable for the problem of interest may significantly improves the efficiency of machine learning\cite{Ramprasad:2017aa}. 
However, such representation omits information about the chemical bonds between the monomers. An alternative is to consider oligmer consisting of $n$ monomers, but there is no clear answer to how to pick $n$. The larger $n$ is, the more representative the oligmer would be to the polymer. Nevertheless, certain types of descriptors, such as the physical descriptors, will take significantly longer time to calculate for a larger molecule. Furthermore, some fingerprints could be biased toward polymers with smaller or larger monomers depending on the choice of $n$. For example, a fingerprint that counts if there are more than 3 benzene rings in a molecule will not be able to distinguish between polystyrene and poly(bisphenol A carbonate) when $n > 3$. Figure \ref{fig:fingerprints} shows how different fingerprints may have different convergence behavior with respect to $n$. \citeauthor{Wu:2016aa} proposed to calculate descriptor of polymers using an infinitely long chain of their corresponding monomers\cite{Wu:2016aa}. However, the bias issue for some fingerprints remains unsolved.

Polymer chain can be separated into a backbone and its side chains. Distinguishing the two components is theoretically important to predict polymer properties. However, it is not clear how to efficiently encode such information into a general descriptor for different classes of polymers. Similarly, developing descriptors for polymers consisting of more than one type of monomer is still an open challenge. For alternating copolymer, we can simply consider the combined repeating unit of multiple monomers as a ``metamonomer", but the molecule may become too big that conventional descriptors or fingerprints are not efficient anymore. Efficient descriptor for block or graft copolymers is yet to be found. Embedding methods using neural network can be a potential option, but problems such as invariance to the input order of monomers or generality to different types of chemical reactions used to form the polymers using multiple monomers remain to have no clear solution.

\section{Predictive models}

One of the key components in polymer design is the prediction of polymer properties relevant to a target application. Some of the important properties often considered in the polymer industry include transparency, glass transition temperature, toughness, etc\cite{Willbourn1976}. The ability to predict properties of different classes of polymers opens up new computational design approaches for polymers, such as high-throughput screening and inverse design, which will be discussed in the next section. One approach to predict polymer properties is building empirical equations based on physical descriptors and some basic properties of polymers. An effort to collect such equations in Python has been made by the \textit{thermo} package\cite{thermo}. This approach can be very powerful, but could be difficult to use when predicting properties for a new design with not much knowledge available yet. Another approach that is more suitable for design purpose, called the quantitative structure–property relationship (QSPR) modeling, aims at mapping structural descriptors of a material to its property. In particular, models that rely only on 2D structure information of the polymer is preferred as it avoids intensive geometry optimization for the polymer molecules. To date, around 300 articles can be found on the web of science database when searching with keywords QSPR and polymer. 

Group contribution method is one of the earliest data-driven approaches to predict complex properties of different polymers\cite{VKBook}. The basic idea is that certain groups of bonded atoms within different molecules may have common effects on a specific material property. Linear regression is used to model the interaction between groups when data is limited, but higher order models can also be used. Machine learning models using fingerprints as input can be seen as an extension to this idea, since the fingerprints are collections of rules that check the existence of different structural groups. Some commonly used models include elastic net, support vector machine, random forest, Gaussian process, neural network, etc. Table \ref{tab:pred} summarizes some recent applications of these algorithms in polymer informatics. 

\begin{table}
  \caption{Examples of QSPR for different polymer properties using machine learning technologies. Corresponding references are cited in the property column. For properties, $\Delta E$ denotes atomization energy, $\epsilon_{gap}$ denotes bandgap, $\kappa$ denotes dielectric constant, $\rho$ denotes density, HOMO denotes highest occupied molecular orbital, LUMO denotes lowest unoccupied molecular orbital, $\epsilon_{opt}$ denotes optical gap, $\eta$ denotes refractive index, $\delta$ denotes solubility parameter, $T_g$ denotes glass transition temperature, $E_g$ denotes glass modulus, $E_r$ denotes rubber modulus, and tan$\delta_{max}$ denotes peak height of viscoelastic loss tangent. For descriptors, \textit{Mix} denotes a mix of various descriptors specified by \citeauthor{Kim:2018aa}\cite{Kim:2018aa}, \textit{ICD} denotes the infinite chain descriptors\cite{Wu:2016aa}, \textit{Str} denotes customized strings by \citeauthor{doi:10.1063/1.5023563}\cite{doi:10.1063/1.5023563}, \textit{D\&P} denotes a combination of the Dragon\cite{Dragon} and PaDEL\cite{Yap:2011aa} descriptors, and \textit{Img} denotes direct use of 2D microstructure images. For models, \textit{GP} denotes Gaussian process, \textit{SVM} denotes support vector machine, \textit{PLS} denotes partial least squares regression, \textit{VAE} denotes using the best regression model based on the hidden layer of a variational autoencoder as described by \citeauthor{doi:10.1063/1.5023563}\cite{doi:10.1063/1.5023563}, and \textit{CNN} denotes a multi-task learning convolutional neural network. For test method to calculate root mean squared error (RMSE), mean absolute error (MAE) and coefficient of determination (R${}^2$) of the QSPR models, \textit{CV-5} denotes a 5-fold cross validation, \textit{Split-X} denotes a X\% random splitting of test data from the full data set, and \textit{Select-27} denotes manually picking 27 data points as test data. (${}^*$~Mean absolute percentage error is measured for these studies)} 
  \label{tab:pred}
  \begin{tabular}{lllllllll}
    \hline
    Property & Data size & Descriptor & Model & Test method & RMSE & MAE & $\text{R}^2$ & Unit \\
    \hline
    $\Delta E$\cite{Kim:2018aa} & 392 & Mix & GP & CV-5 & 0.01 & 0.01 & 0.999 & eV/atom\\
    $\epsilon_{gap}$\cite{Wu:2016aa} & 155 & ICD & SVM & Split-20 & --- & --- & 0.88 & eV \\
    $\epsilon_{gap}$\cite{Kim:2018aa} & 382 & Mix & GP & CV-5 & 0.3 & 0.23 & 0.971 & eV \\
    $\epsilon_{gap}$\cite{doi:10.1063/1.5023563} & 3,989 & Str & VAE & CV-5 & --- & 74 & --- & meV\\
    $\kappa$\cite{Wu:2016aa} & 155 & ICD & SVM & Split-20 & --- & --- & 0.96 & --- \\
    $\kappa$\cite{Kim:2018aa} & 384 & Mix & GP & CV-5 & 0.48 & 0.32 & 0.815 & --- \\
    $\rho$\cite{Kim:2018aa} & 173 & Mix & GP & CV-5 & 0.05 & 0.03 & 0.938 & g/cm${}^3$ \\
    HOMO\cite{doi:10.1063/1.5023563} & 3,989 & Str & VAE & CV-5 & --- & 66 & --- & meV\\
    LUMO\cite{doi:10.1063/1.5023563} & 3,989 & Str & VAE & CV-5 & --- & 43 & --- & meV\\
    $\epsilon_{opt}$\cite{doi:10.1063/1.5023563} & 3,989 & Str & VAE & CV-5 & --- & 70 & --- & meV\\
    $\eta$\cite{Kim:2018aa} & 384 & Mix & GP & CV-5 & 0.08 & 0.05 & 0.892 & --- \\
    $\eta$\cite{Khan:2018aa} & 221 & D\&P & PLS & Split-30 & --- & 0.004 & 0.899 & --- \\
    $\eta$\cite{doi:10.1063/5.0008026} & 527 & Mix & GP & Select-27 & 0.05 & --- & 0.88 & --- \\
    $\delta$\cite{Kim:2018aa} & 113 & Mix & GP & CV-5 & 0.56 & 0.4 & 0.955 & MPa${}^{1/2}$ \\
    $T_g$\cite{Wu:2016aa} & 270 & ICD & SVM & Split-20 & --- & --- & 0.95 & K \\
    $T_g$\cite{Kim:2018aa} & 451 & Mix & GP & CV-5 & 17.74 & 12.79 & 0.944 & K \\
    $E_g$\cite{D0ME00020E} & 11,000 & Img & CNN & Split-15 & --- & 0.68 & --- & \%${}^*$ \\
    $E_r$\cite{D0ME00020E} & 11,000 & Img & CNN & Split-15 & --- & 3.12 & --- & \%${}^*$ \\
    tan$\delta_{max}$\cite{D0ME00020E} & 11,000 & Img & CNN & Split-15 & --- & 3.58 & --- & \%${}^*$ \\
    \hline
  \end{tabular}
\end{table}

\begin{figure}
  \includegraphics[trim={0 0 0 0},clip, width=16cm]{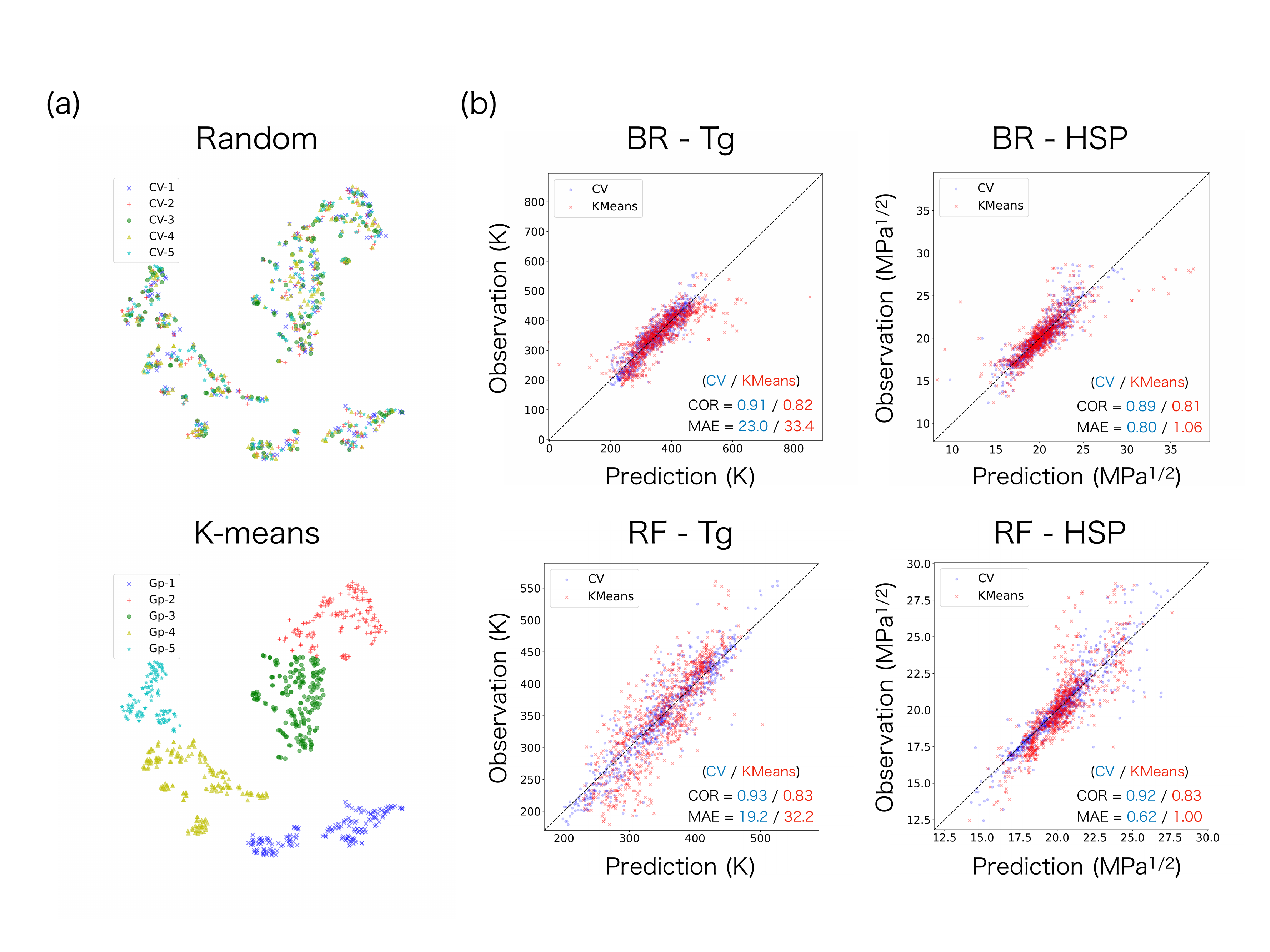}
  \caption{Demonstrating extrapolation power of machine learning models using data from Polymer Genome\cite{Kim:2018aa}. (a) Visualization of a 2D projection of the 200 physical descriptors in RDKit using t-distributed Stochastic Neighbor Embedding\cite{TSNE} with perplexity = 30. The data is further separated into five groups, either randomly or through K-means clustering\citep{kmeans} with number of cluster = 5. The groupings are used for cross validation to test the prediction performance on glass transition temperatures (Tg) and Hildebrand solubility parameter (HSP). (b) Plots of predictions versus observed data for Tg and HSP based on Bayesian ridge regression (BR) and random forest (RF). Crosses are results from cross validation based on K-means clustering, i.e., each set of test data is not similar to the training data (extrapolation). Circles are results from cross validation based on random picking. The extrapolation case has a worse prediction accuracy all four cases, where the failure patterns are different between BR and RF.}
  \label{fig:P2O}
\end{figure}

Machine learning models are inherently interpolative, i.e., their predictions are reliable only around the domain close to the training data. The range of properties prediction for certain types of polymers with a reasonable accuracy governs the potential search space one can cover, i.e., the performance of the final design. In other words, the training data determines the feasible design space of the target application. The concept of applicability domain (AD) developed in Cheminformatics is used to quantify the reliable region of a QSPR model.\cite{Sheridan:2004aa} The concept of uncertainty in statistics is also a popular metric believed to be strongly correlated with the validity of a prediction.\cite{Chatfield:1995aa} Figure \ref{fig:P2O} shows how different models fail to predict different materials properties when extrapolating from the given data. One idea to tackle this issue is called transfer learning, that is to exploit information learned from a relevant task for improving prediction of another task. \citeauthor{Yamada:2019aa} demonstrated the successful applications of transfer learning in different materials science problem, including polymers\cite{Yamada:2019aa}. Intuitive ideas for knowledge transfer include transferring from a global material space to a local domain, from a material property with rich data to a physically linked property with little data, or from computational data to experimental data. The latter idea is also called multi-fidelity learning, which has been successfully applied to predict crystallization tendency\cite{Venkatram:2020aa} and bandgap\cite{PATRA2020109286} of polymers. 

\section{Polymer design}

While there are increasing examples of machine-assisted polymer design, there has not been any report of end-to-end design example that covers every step from monomer design to manufacturing process. Instead, polymer informatics has been used to improve design efficiency within each step of of the polymer design process. For example, \citeauthor{Wu:2019aa} discovered new homopolymers with high thermal conductivity that are validated experimentally\cite{Wu:2019aa} and \citeauthor{C7ME00131B} developed an algorithm that discovers optimal strategy to experimentally control MWDs of various polymers\cite{C7ME00131B}. There are three types of design strategies based on machine learning technology: high-throughput screening, inverse design, and experimental design. We will discuss various efforts of applying these methods to polymer design in this section. 

\subsection{High-throughput screening}
High-throughput screening aims at identifying potential candidates of interest by conducting computational or experimental tests on a large pool of candidates. In polymer informatics, a library of chemically or synthetically feasible polymer candidates is built and then computationally screened by predictive models relevant to the target material properties. When the search space is finite and tractable, the library is simply composed of the exhaustive list of candidates, such as selection of optimal processing method within the existing technologies. One can also use a library from any open databases. Otherwise, a generative algorithm is needed to construct a pool of candidates of interest. For example, a conventional approach is to define a set of molecular fragments of interest and exhaustively combine different numbers of fragments up to an upper limit to form the candidate set. Fragment can be obtained from organic molecules which have more large open databases, such as GDB-17\cite{Ruddigkeit:2012aa} or Pubchem\cite{Kim:2016}. To expand beyond the search space limited by a finite variety of fragments, \citeauthor{doi:10.1002/minf.201900107} implemented a language model proposed by \citeauthor{Ikebata:2017}\cite{Ikebata:2017} on polymers represented in SMILES.\cite{doi:10.1002/minf.201900107} Different deep neural networks are also used to learn the SMILES or graph representation of organic molecules\cite{Cao2018MolGANAI, 10.5555/3327345.3327537, Popovaeaap7885}.

Giving a library of candidates, one can use pretrained predictive models to computationally screen out candidates with target properties. There is a significant amount of literature that applies high-throughput screening to different polymer applications. Recent examples of high-throughput screening in polymer designs using machine learning models include searching of high refractive index polymers\cite{Khan:2018aa, JABEEN2017215, Afzal:2019aa} and screening optoelectronic properties of conjugated polymers\cite{Wilbraham:2018aa}. This strategy of polymer design may sound inefficient to solve a single design problem, because a significantly large library is needed to increase the probability of finding candidates of interest. However, a well-developed library that contains diverse candidates can be reused for many different design problems. With a rich database of candidates, high-throughput screening is a very attractive tool in many industrial applications. 

\subsection{Inverse design}
Another design strategy, referred as the inverse design, is to perform targeted search in a materials space guided by knowledge extracted from existing data. While predictive model is a mapping from an input $x$ (polymer) to an output $y$ (material property), the goal of inverse design is to map a targeted range of $y$ to a sub-domain of $x$. This can be achieved by solving an optimization problem using sophisticated algorithms, such as genetic algorithms, or by sampling the set of $x$ with high probability to be in the targeted range of $y$ under a Bayesian framework. Both approaches can be implemented in an iterative algorithm:
\begin{enumerate}
    \item Start with a set of initial candidates.
    \item Propose a new set of candidates based on the existing ones.
    \item Evaluate likeliness of the candidates to have the desired material properties.
    \item Repeat step 2 and 3 for fixed number of times or until convergence.
\end{enumerate}
In step 2, a generative algorithm is needed to propose new candidates similar to the one in the high-throughput screening. In the case of optimization, the likeliness metric in step 3 is usually a loss function measuring the distance of the predicted properties for the candidates and the targeted properties. For sampling, the likeliness metric is usually correlated with the probability of observing the candidates conditional on the targeted properties. 

The inverse design problem is inherently ill-posed, i.e., there are different types of materials that may have the target properties. It is important that an inverse design algorithm can produce a diverse set of candidates in order to maximize the probability of discovering novel functional materials. Typical optimization algorithm can only search for the optimal solution. Sampling methods can have a higher chance to provide diverse candidates, but can also be trapped in a local mode, especial when the search space is high dimensional. Many efforts have been made to address this issue, such as limiting the search space size, projecting the search space to a low dimensional space, or employing an annealing algorithm, etc. Table \ref{tab:inv} shows some recent examples of successful inverse design of different polymer applications using machine learning technologies. 

\begin{table}
  \caption{Examples of polymer inverse design using machine learning technologies.}
  \label{tab:inv}
  \begin{tabular}{lll}
    \hline
    Publication & Target property & Method \\
    \hline
    Mannodi-Kanakkithodi & Dielectric constant & Optimization using \\
    et al. (\citeyear{Mannodi-Kanakkithodi:2016aa})\cite{Mannodi-Kanakkithodi:2016aa} & and bandgap & genetic algorithm \\[0.3cm]
    \multirow{2}{*}{\citeauthor{doi:10.1063/1.5023563} (\citeyear{doi:10.1063/1.5023563})\cite{doi:10.1063/1.5023563}} & LUMO and & Gradient-based optimization on \\
    & optical gap energy & embedded space in deep neural network\\[0.3cm]
    \multirow{2}{*}{\citeauthor{Pilania:2019aa} (\citeyear{Pilania:2019aa})\cite{Pilania:2019aa}} & Glass transition & Optimization using \\
    & temperature & genetic algorithm \\[0.3cm]
    \multirow{2}{*}{\citeauthor{Kumar:2019aa} (\citeyear{Kumar:2019aa})\cite{Kumar:2019aa}} & \multirow{2}{*}{Phase behavior} & Optimization using \\
    & & particle swarm optimization \\[0.3cm]
    \multirow{2}{*}{\citeauthor{Wu:2019aa} (\citeyear{Wu:2019aa})\cite{Wu:2019aa}} & Thermal & Sampling with \\
    & conductivity & sequential Monte Carlo \\[0.3cm]
    \multirow{2}{*}{\citeauthor{Schadler_2020} (\citeyear{Schadler_2020})\cite{Schadler_2020}} & Three different & Optimization using \\
    & dielectric properties & genetic algorithm \\[0.3cm]
    \multirow{2}{*}{\citeauthor{doi:10.1002/minf.201900107} (\citeyear{doi:10.1002/minf.201900107})\cite{doi:10.1002/minf.201900107}} & Dielectric constant & Sampling with \\
    & and bandgap & sequential Monte Carlo \\
    \hline
  \end{tabular}
\end{table}

\subsection{Experimental design}
The previous two design strategies relies heavily on efficiency of the generative model and accuracy of the predictive model, which, in turn, relies on the quality and quantity of training data. Since many properties of polymers have only limited data, increasing data size with extra experimental or computational tests is inevitable to ensure that the machine learning models can cover a large enough design space. Instead of randomly picking candidates to perform the new tests, one can employ a recursive design process, i.e., new test candidates are optimally selected and tested, and the results are added to the existing data set for improving the optimal candidate selection in the next round of tests. Such trial-and-error design process is what a chemist would do in practice. 

The goal of experimental design is to minimize the amount of new experiments required to reach a design goal. There are two perspectives to the optimality of candidate selection: (1) exploit the knowledge embedded in the existing data to make the best guess of which candidates may satisfy the design goal, or (2) explore candidates with the least information from the existing data to infer their properties. In a typical experimental design algorithm, we define a utility function that balance the tradeoff between the two perspectives. Candidates that optimize the utility function are selected for the next round of tests. Bayesian optimization and reinforcement learning are two commonly used algorithms for experimental design. The former considers candidates with a high prediction uncertainty as exploratory and optimize the utility function accordingly. The latter treats the problem as a game with reward when achieving the design goal, where an agent is trying to learn the best strategy (minimum number of experiments) to get the maximum reward (reaching the design goal). Table \ref{tab:exp} shows some recent examples of experimental design applications on different stages of the hierarchical polymers design process.

\begin{table}
  \caption{Examples of polymer experimental design using machine learning technologies.}
  \label{tab:exp}
  \begin{tabular}{llll}
    \hline
    Publication & Target & Search space & Method \\
    \hline
    \multirow{2}{*}{\citeauthor{Li:2017aa} (\citeyear{Li:2017aa})\cite{Li:2017aa}} & Median length, median & Five synthetic & Bayesian \\
    & diameter and quality of fibers & process parameters & optimization \\[0.3cm]
    \multirow{2}{*}{\citeauthor{C7ME00131B} (\citeyear{C7ME00131B})\cite{C7ME00131B}} & \multirow{2}{*}{Shape of MWD} & Amount of five & Reinforcement\\
     & & chemical reagents & learning \\[0.3cm]
    \multirow{2}{*}{\citeauthor{WANG2018146} (\citeyear{WANG2018146})\cite{WANG2018146}} & Interphase properties & Hyperparameters in & Bayesian \\
    & (dielectric and viscoelastic) & two interphase models & optimization \\[0.3cm]
    \multirow{2}{*}{\citeauthor{minami_kawata_fujita_murofushi_uchida_omori_okuno_2019} (\citeyear{minami_kawata_fujita_murofushi_uchida_omori_okuno_2019})\cite{minami_kawata_fujita_murofushi_uchida_omori_okuno_2019}} & Glass transition & Mixing ratios of & Bayesian \\
    & temperature & three selected polymers & optimization \\[0.3cm]
    \multirow{2}{*}{\citeauthor{kim_chandrasekaran_jha_ramprasad_2019} (\citeyear{kim_chandrasekaran_jha_ramprasad_2019})\cite{kim_chandrasekaran_jha_ramprasad_2019}} & Glass transition & 736 predefined & Bayesian \\
    & temperature & candidates in database & optimization \\
    \hline
  \end{tabular}
\end{table}

Experimental design algorithm is the ideal solution when we do not have a large enough polymer database to begin with. However, polymer experiments are costly and syntheses of new polymers are difficult. Computational simulations are becoming more and more accessible, yet calculations for new polymers often required labor-intensive tuning of model parameters. Automatic simulation, synthetic planning, and property measurement for polymers remain challenging and are continuously explored by researchers in polymer informatics.

\section{Discussion}

Polymer informatics is a promising tool for discovery of novel polymers. With a sufficiently large data set to support the use of modern machine learning technology, a data-driven approach of polymer design will significantly improve the pace of making new functional polymers, satisfying the rapidly expanding demand on polymeric materials in modern society (e.g., deformable electronic devices\cite{MCBRIDE202072}). One of the most important elements of polymer informatics is the availability of large open databases. Building such databases for polymer is challenging because of many reasons: (1) difficulty of encoding the hierarchical structure of polymers for machine learning purposes, (2) inconsistent naming rules throughout the history of polymer science, (3) lack of data sharing due to many privately own industrial data, etc. Nonetheless, this is an essential step towards a fully data-driven design process. Many efforts have been made to build the backbone technologies necessary to exploit the potential benefit of polymer informatics, such as developing new descriptors to better capture the physical properties of polymers and new simulation methods to estimate polymer properties with higher accuracy in a shorter time. The true power of polymer informatics is to release polymer scientists from the low efficiency trial-and-error design process, thus, to free up more time for higher level design concepts and theoretical advancements. Such opportunity can be realized only if we work together to push for a more open community in polymer science, where everyone can benefit from the new paradigm of polymer design. An easy first step to take would be to engage in the field of polymer informatics and experience the new way of studying polymers ourselves.



\begin{acknowledgement}

This work was supported in part by the “Materials Research by Information Integration” Initiative (MI2I) project of the Support Program for Starting Up Innovation Hub from Japan Science and Technology Agency (JST). S.W. gratefully acknowledges financial support from JSPS KAKENHI Grant Number JP18K18017. R.Y. acknowledges financial support from a Grant-in-Aid for Scientific Research (B) 15H02672 and a Grant-in-Aid for Scientific Research (A) 19H01132 from the Japan Society for the Promotion of Science (JSPS).

\end{acknowledgement}


\bibliography{PI_ISMSI2021}

\end{document}